# TEMPERATURE AND DENSITY EFFECTS ON THE LOCAL SEGMENTAL AND GLOBAL CHAIN DYNAMICS OF POLY(OXYBUTYLENE)

R. Casalini[†,§] and C.M. Roland[†]

[†] Naval Research Laboratory, Chemistry Division, Code 6120, Washington DC 20375-5342
[§] George Mason University, Chemistry Dept., Fairfax VA  22030

**ABSTRACT** Dielectric spectroscopy measurements over a broad range of temperature and pressure were carried out on poly(oxybutylene) (POB), a type-A polymer (dielectrically-active normal mode). There are three dynamic processes appearing at lower frequency, the normal and segmental relaxation modes, and a conductivity arising from ionic impurities. In combination with pressure-volume-temperature measurements, the dielectric data were used to assess the respective roles of thermal energy and density in controlling the relaxation times and their variation with T and P. We find that the local segmental and the global relaxation times are both a single function of the product of the temperature times the specific volume, with the latter raised to the power of 2.65. The fact that this scaling exponent is the same for both modes indicates they are governed by the same local friction coefficient, an idea common to most models of polymer dynamics. Nevertheless, near $T_g$, their temperature dependences diverge. The magnitude of the scaling exponent reflects the relatively weak effect of density on the relaxation times. This is usual for polymers, as the intramolecular bonding, and thus interactions between directly bonded segments, are only weakly sensitive to pressure. This insensitivity also means that the chain end-to-end distance is invariant to P, conferring a near pressure-independence of the (density-normalized) normal mode dielectric strength.

The ionic conductivity dominates the low frequency portion of the spectra. At lower temperatures and higher pressures, this conductivity becomes decoupled from the relaxation modes (different T-dependence), and exhibits a significantly weaker density effect. At frequencies higher than the structural relaxation, both an excess wing on the flank of the α-peak and a secondary relaxation are observed. From their relative sensitivities to pressure, we ascribe the former to an unresolved Johari-Goldstein (JG) relaxation, while the higher frequency peak is unrelated to the glass transition. These designations are consistent with the relaxation time calculated for the JG process.

The dynamic properties of the POB are essentially the same as those of polypropylene glycol, in accord with their similar chemical structures. However, POB is less fragile (weaker $T_g$-normalized temperature dependence), its relaxation times less sensitive to density changes, and facilitating the measurements herein, its normal mode has a substantially larger dielectric strength

**Introduction**

The dynamics of glass-forming liquids and polymers remains a challenging problem, not only in relating structure to properties, but in simply understanding whether and how the various relaxation processes relate to the glass transition. Although relaxation in the equilibrium state just above the glass transition temperature, $T_g$, receives the bulk of attention, the relationship to $T_g$ of secondary processes, in particular the Johari-Goldstein relaxation[1,2,3], the "nearly constant loss" [4,5,6,7,8], and even the Boson peak[9] have been explored by various groups. In addressing this problem satisfactorily, it is necessary to use additional experimental parameters, for example, pressure[10,11,12,13,14,15,16,17], nano-scale confinement[18,19,20,21], and crosslinking [22,23,24,25,26,27,28]. A recent interest is how global motions of polymer chains relate to their local dynamics. This can be investigated using diffusion measurements[29] and, over a limited frequency range, by mechanical spectroscopy[30,31]. For the (relatively few) polymers having a dipole moment parallel to their chain backbone ("type A" polymers[32]), the global motions can be followed dielectrically, along with the "α-relaxation" ($T_g$ dynamics). Certainly for studying the α-relaxation, dielectric spectroscopy is one of the more powerful tools, able to non-intrusively probe molecular motions over a broad frequency range (more than 12 decades).[33]

A central issue concerning the dynamics in condensed matter is what causes the spectacular change in relaxation time, τ, as $T_g$ is approached. Proposed mechanisms abound: (i) jamming (spatial constraints) of molecular motions due the increasing density (i.e. reduced free volume) [34,35,36], (ii) progressive trapping of molecules or polymer segments within potential wells[37], (iii) a reduction in the number of available configurations (entropy loss)[38], and (iv) greater intermolecular cooperativity of the molecular motions [39,40]. There are two main questions – is there a dominant thermodynamic variable which controls vitrification, and what is the connection between the thermodynamics and dynamics. By combining spectroscopy measurements with the equation of state, the relative degree to which thermal energy and density govern the temperature dependence of τ can be quantified. While analysis of a limited number of materials led to the conclusion that thermal energy is the dominant variable controlling τ(T),[41] more recent studies of a broader number of glass formers clearly show that both temperature and density play significant roles. Their relative contribution to glass-formation varies with chemical structure, and for polymers density seems to exert less of an influence than for molecular glass-formers[42,43,44]. This is ironic, given the historic prominence of free volume theories in the study of polymer dynamics. Recently, we have extended this type of analysis by demonstrating that local (dielectric α) relaxation times, measured under various conditions of T and P, can be superposed when plotted versus the product of the temperature times the specific volume, with the latter quantity raised to a material-specific exponent, γ [45,46]. The magnitude of this exponent reflects the relative contribution of volume (density) to the dynamics. Although determined empirically from superpositioning of τ(T,P) data, for both non-associated liquids and polymers the exponent can be related to the repulsive part of the intermolecular potential[47,48,49]. Recalling the classic van der Waals model, the idea is that for very local processes, the attractive part of the potential is averaged (thus imposing only an isotropic pressure), whereby the details of the repulsive potential govern the local properties.[50,51,52]. The interpretation is that segmental motion is thermally-activated, with the height of the potential barriers dependent on the density. Thus, the assumption of an inverse power law for the repulsive potential[49,50,53,54], gives rise to a particular scaling of the local relaxation times[47,48]. In the case of polymers, the values of the exponent γ are often rather small, suggesting an intermolecular potential which departs from the typical Lennard-Jones form. It is likely that intramolecular cooperativity[55] (due to connectivity of the chain units) strongly influences the effective local potential. A similar situation may pertain for hydrogen-bonded materials.

    Our interests herein are the normal mode motions and their relationship to the local segmental dynamics. We have recently shown that for two type-A polymers, polypropylene glycol (PPG) and 1,4-polyisoprene (PI), the normal mode relaxation times (strictly speaking, the longest normal mode relaxation times) superpose onto a single master curve, using the same value of the scaling exponent as for the segmental relaxation times. While this scaling itself is provocative, the equivalence of the scaling parameter for the two processes suggests that the factors underlying the dynamics are the same. This, in fact, is a central assumption underlying

the Rouse and reptation models of polymer dynamics – that the chain modes are governed by the same local friction coefficient associated with segmental motion[34,56].

In this work we investigate polyoxybutylene (POB, or poly(1,2-butylene oxide)), a type-A polymer. Apart for the pendant ethyl group in place of the methyl group, POB has the same chemical structure as PPG (their structures are depicted in Figure 1). PPG has been the subject of many studies, by dielectric spectroscopy[10,14,15,57,58,59,60,61,62,63] and other techniques[64,65,66,67]. Kyritsis et al.[68] recently reported dielectric measurements at ambient pressure on several POB samples, as well as POB block copolymers. Herein, we focus on a single POB of *ca.* 5,000 D molecular weight. We have carried out broad band dielectric relaxation and specific volume measurements, both as a function of temperature and pressure.

From a comparison of the dielectric spectra of PPG and POB having the same degree of polymerization (Figure 1), it is evident that the POB normal mode has a much larger dielectric strength. For a monodisperse type A polymer, the dielectric strength of the normal mode $\Delta\varepsilon_N$ is [69,70]

$$\Delta\varepsilon_N = \frac{4\pi F \tilde{\mu}_p^2 N_A \rho}{3 k_B T M_w} \langle r^2 \rangle \qquad (1)$$

where $\tilde{\mu}_P$ is the dipolar moment per unit length of the repeat unit parallel to the backbone, $N_A$ is Avogadro's number, $M_w$ the molecular weight, $\rho$ the density, $k_B$ the Boltzmann constant, $\langle r^2 \rangle$ the mean square end-to-end distance of the chain, and $F$ is the local field correction (F ~ unity[70]). Therefore, for POB and PPG of equivalent chain lengths and density, and assuming that $\tilde{\mu}_P$ is the same, the larger $\Delta\varepsilon_N$ of POB is ascribed to a larger $\langle r^2 \rangle$. This results from the larger pendant moiety, and consequently less flexible chain. The greater strength of the normal mode for POB facilitates dielectric measurements, an especially important consideration for high pressure measurements. A large $\Delta\varepsilon_N$ also makes more accurate deconvolution of the normal mode from the nearby segmental relaxation.

It is of interest to see if other properties reflect this apparent difference in chain flexibility. While the dynamics of the normal mode is independent of local chemical structure, segmental relaxation, arising from dipoles perpendicular to the chain (type B dipoles), depends strongly on the details of the local structure; e.g., barrier heights to internal rotation, energy differences between rotational isomeric states, the local friction coefficient, and steric constraints on local motions[71,72]. Notwithstanding these possible differences between POB and PPG, it is interesting to note that the shapes of the α-relaxation peaks in Figure 1 are the same for the two polymers.

In this work, by combining dielectric data with PVT measurements, we are able to analyze the factors governing both the global and segmental dynamics. We also examine ionic conductivity in the POB, in order to assess the degree of enhancement of translational motions, relative to the reorientational relaxation. Finally, we describe observations at high frequencies of both an excess wing and a secondary dispersion in POB.

**Experimental**

The POB was from Mark Nace of Dow Chemical Co., courtesy of Colin Booth of the University of Manchester. The sample had $M_n$=4800 and $M_w/M_n$=1.10. This corresponds to 67 repeat units per chain, about equal to that of the PPG (molecular weight = 4000 g/mol) studied previously[14].

Dielectric spectra were obtained with a parallel plate geometry using an IMASS time domain dielectric analyzer ($10^{-4}$ to $10^3$ Hz) and a Novocontrol Alpha Analyzer ($10^{-2}$ to $10^6$ Hz). For measurements at elevated pressure, the sample was contained in a Manganin cell (Harwood Engineering), with pressure applied using a hydraulic pump (Enerpac) in combination with a pressure intensifier (Harwood Engineering). Pressures were measured with a Sensotec tensometric transducer (resolution = 150 kPa). The sample assembly was contained in a Tenney Jr. temperature chamber, with control to within ± 0.1 K at the sample.

PVT measurements employed a GNOMIX apparatus, modified to allow sub-ambient temperatures. Typically, the pressure was varied from 10 to 200 MPa at fixed temperatures ranging from *ca.* -10 to 95 ºC.

**Results and Discussion**

**Shape of the relaxation functions.** Dielectric spectra were obtained at atmospheric pressure at various temperatures, and over a range of pressures at each of three temperatures. A representative loss curve measured at 246.6 K is shown in Figure 2, revealing three contributions to the dielectric response. At lower frequencies, there is a contribution to the dielectric loss, due to conduction (charge transport) of ionic impurities. Observed at successively higher frequencies are the normal mode peak, reflecting motion of the chain end-to-end vector, and the α-relaxation peak, due to local segmental motions. There is some overlap of these processes, so to extract relaxation times the spectra were fitted to the sum of a conductivity term, given by $\sigma_{DC}\omega^{-1}$, where $\sigma_{DC}$ is the conductivity and ω angular frequency, and two Havriliak-Negami (HN) functions[73],

$$\varepsilon" = \Delta\varepsilon \left[1 + 2(\omega\tau_{HN})^\alpha \cos(\pi\alpha/2) + (\omega\tau_{HN})^{2\alpha}\right]^{-\beta/2} \sin(\beta\psi) \qquad (2)$$

where

$$\psi = \arctan\frac{\sin(\pi\alpha/2)}{(\omega\tau_{HN})^{-\alpha} \cos(\pi\alpha/2)} \qquad (2a)$$

In this expression, $\tau_{HN}$ is a relaxation time, and *α* and *β* are shape parameters, describing respectively the symmetrical and unsymmetrical broadening of the dispersion. To minimize the number of free parameters, we fit the spectra for which the segmental and normal mode peaks were well separated, and then used the obtained α (= 0.94 and 0.85 for the normal and segmental modes, respectively) in fitting all other spectra. The three distinct contributions to the spectrum are shown in Figure 2, although excluded from the fitting is the weak peak observed at the highest frequencies. Discussed below, this relaxation process becomes more prominent in measurements at higher pressures and lower temperatures.

In Figure 3 the dielectric strengths determined from fitting the experimental spectra are plotted, after normalization by the density, as a function of the normal mode relaxation time (this serves as a basis to compare the data obtained at various temperatures and pressures). At constant pressure, both modes increase in intensity with decreasing temperature (i.e., with larger $\tau_N$). This is due primarily in the case of the normal mode to an increase of $\langle r^2 \rangle$, and for the segmental to an enhanced correlations among the dipoles.[73] However, at a given temperature, pressure has a negligible effect on the dielectric strength of the normal mode, in agreement with the negligible effect of pressure on $\langle r^2 \rangle$.

In contrast, the dielectric strength for the segmental process increases with pressure, $d\Delta\varepsilon_\alpha/dP \sim 1$ GPa$^{-1}$. The relevant equation is the Kirkwood-Frölich relation [73,74]

$$\Delta\varepsilon_\alpha = g\frac{4\pi F \tilde{\mu}^2 x N_A \rho}{3k_B T M_w} \qquad (3)$$

where $\tilde{\mu}$ is the dipole moment of the repeat unit and $x$ is the degree of polymerization ($xN_A\rho/M_w$ is the number density of repeat units). The Kirkwood/Fröhlich correlation factor, $g$, measures the correlation among dipole moments in neighboring segments. The observed increase of $\Delta\varepsilon_\alpha/\rho$ at constant $T$ seems to indicate that with increasing $P$, there is increased correlation of the orientation among dipoles. Interestingly from Figure 3, the values of $\Delta\varepsilon_\alpha/\rho$ are very close when compared at the same relaxation time. This indicates that for the same value of $\tau$, there is a fixed degree of correlation among dipoles, independent of the particular combination of T and P. Similar behavior was found for the structurally-related PPG[14]. These trends mean that the magnitude of $\Delta\varepsilon_\alpha/\Delta\varepsilon_N$ varies in opposite fashion to that found for PI, a non-associated polymer[12]. The notable difference between the two materials is the stronger correlation effects for polymers such as POB, which have the capacity for hydrogen bonding of the chain ends.

Figure 4 shows measurements at four conditions of T and P, chosen such that the segmental relaxation times are about the same, $\tau_\alpha \sim 1$ ms. Normalizing the spectra by the intensity of the segmental peak makes clear the increase of its relative intensity with either temperature and pressure. It is also apparent from Figure 4 that at this fixed value of the relaxation time, the shape of the relaxation peak for either mode is constant. In fact, this superpositioning of peaks for fixed $\tau$ is observed over the entire range of the measurements. As shown in Figure 5, the product of the two shape parameters, α and β in eq. 2, shows little variation with changes in either T or P, and is constant when compared at a given value of the relaxation time. This quantity, $\alpha\beta$, is a measure of the breadth of the relaxation peak. The modest increase in $\alpha\beta$ for the segmental mode at shorter values of $\tau_N$ may, at least to some extent, reflect an increasing contribution of (unresolved) secondary processes. For $\tau_N > 10^{-2}$ s, $\alpha\beta = 0.43 \pm 0.1$ for the segmental mode, and $\alpha\beta = 0.61 \pm 0.2$ for the normal mode. In terms of the Kohlrausch-William-Watts stretch exponent, $\beta_{KWW} = 0.50$ (using the formula[75] $\alpha\beta = \beta_{KWW}^{1.23}$) for the peak; direct fitting of the data to the transform of the KWW function, which emphasizes the central part of the peak, yields $\beta_{KWW} = 0.52$.

The variation of the shape parameters in Figure 5 is about the same for the two relaxation processes. While the shape of the normal mode peak reflects in part the sample's polydispersity ($M_w/M_n = 1.10$), the breadth of the segmental mode depends on the chemical structure of the repeat unit[71]. It is expected empirically[76] and on theoretical grounds[39,40] that the sensitivity of relaxation times to temperature will correlate with the breadth of the dispersion. Thus, the inference from Figure 5 is that the separation of the normal and segmental peaks should not vary strongly with T or P. To probe this in more detail, we analyze the temperature and pressure dependences of the two relaxation times.

**Relaxation times and ionic conductivity.** Segmental and normal mode relaxation times at ambient pressure are shown in Figure 6. These are defined as the reciprocal of the frequency of the maximum in the respective loss peak, which is related to the HN relaxation time according to[73]

$$\tau = \tau_{HN} \left( \sin \frac{\alpha \pi}{2+2\beta} \right)^{-1/\alpha} \left( \sin \frac{\alpha \beta \pi}{2+2\beta} \right)^{1/\alpha} \quad (4)$$

Also included in Figure 6 are relaxation times for a POB having a somewhat larger $M_w$ = 5,085 D [68]. There is good agreement between the two data sets. The results herein were fit to the Vogel-Fulcher-Tammann-Hesse equation[34],

$$\tau(T) = \tau_0 \exp\left(\frac{B}{T-T_0}\right) \quad (5)$$

with the obtained parameters listed in Table 1. Using eq. 5 to interpolate, the dynamic glass transition temperature, defined as $\tau_\alpha(T_g)$ = 100 s, equals 199.0 K at atmospheric pressure. In the inset to Figure 6, the segmental relaxation times are plotted as a function of the $T_g$-normalized temperature. This yields for the fragility, defined as the slope at $T_g$, $\left.\frac{d \log(\tau)}{d(T_g/T)}\right|_{T=T_g} = 73$. Thus, like PPG, POB is a moderately fragile polymer.[76]

The curves for the two relaxation times in Figure 6 are parallel at higher temperatures, but as $T_g$ is approached, the segmental relaxation mode exhibits a stronger variation with temperature. This is a general trend, especially for lower molecular weight polymers, as shown previously for PPG[77,78], PI[79] and polylactide (PLA)[80]. The pressure dependences of the two modes, however, are essentially equivalent. We show this with a double logarithmic plot in Figure 7 of the segmental relaxation times versus the normal mode relaxation times; the slope (log$\tau_\alpha$/log$\tau_N$) is close to unity. Actually, in the vicinity of $T_g$ the normal mode exhibits a slightly stronger pressure dependence, as seen from the plot of the logarithm of the ratio $\tau_N/\tau_\alpha$ in the inset to Figure 7. Analogous to the behavior when temperature is varied, for small values of log($\tau_N$), the quantity log($\tau_N/\tau_\alpha$) is almost constant (i.e., the two modes have an identical pressure dependence), while very close to $T_g$, log($\tau_N/\tau_\alpha$) decreases (i.e. the two relaxations get closer). Commonly, pressure-dependences are expressed in terms of an activation volume, defined as $\Delta V = 2.303 RT \frac{d \log \tau}{dP}$. $\Delta V$ increases with increasing pressure, with the limiting low pressure values listed in Table 2 for the three measured isotherms. As expected from the results in Figure 7, there is no significant difference between the activation volumes for the two relaxation modes, when compared at equal (low) pressure. This equivalence of the pressure-sensitivity of $\tau_\alpha$ and $\tau_N$ is unusual. Previous studies of PI [12], PLA[80] and PPG[14] found the segmental mode to be more sensitive to pressure (although from measurements over a limited range on PPG, the opposite result, $\Delta V_\alpha < \Delta V_N$, was also reported[15]). However, such results depend upon the basis for comparison. For example, the activation volume for the segmental mode in the low P limit can be compared to that of the normal mode, with the latter taken over the same dynamic range (same values of $\tau$), rather than same range of P. This leads to a different conclusion.

In Figures 8 and 9 respectively, we compare the coupling of the two modes over the same P range using published data for PPG[14] and for two different molecular weights of PI[12,81]. In each case, the slope of the curve is very close to unity, indicating that in contrast with previous conclusions, the pressure dependences of the two modes are essentially the same (i.e., $\Delta V_\alpha \sim \Delta V_N$), at least when compared over the same range of P. However, as discussed above, this coupling may break down very close to $T_g$.

The ionic conductivity responds to changes in T and P in a qualitatively similar fashion as the inverse of the segmental relaxation times. However, it is well-known that near $T_g$, translations are often enhanced relative to reorientational motions (such as associated with the dielectric α-process)[82,83,84]. The result is that a double logarithmic plot of σ versus $\tau_\alpha$ will have a slope deviating from unity[85]. This decoupling has been ascribed to spatially heterogeneous dynamics[86] or to differences inherent to the cooperative dynamics of condensed matter[87]. In Figure 10, the conductivities determined for POB at both ambient and elevated pressure are plotted versus $\tau_\alpha$. For all conditions, the power law slope is less than unity (average value = 0.85 ± 0.07), indicating a significant enhancement of ion translation relative to the rate of local segmental motion.

The pressure coefficient of $T_g$ cannot be determined accurately, when the latter is defined as $\tau_\alpha(T_g) = 100$ s. For the temperature at which $\tau_\alpha = 1$ s, $T_\alpha$, we find that $dT_\alpha/dP = 155 \pm 25$ KGPa$^{-1}$. This is significantly larger than the pressure coefficient of more strongly hydrogen-bonded glass-formers, for which values less than *ca.* 90 KGPa$^{-1}$ are typical[16,88]. Non-associated materials have larger pressure coefficients. For a PPG having the same degree of polymerization as the POB herein (x = 67), and thus the same mole fraction of H-bonded repeat units, $dT_\alpha/dP$ equals $177 \pm 36$ GPa$^{-1}$ [14]. Thus, the difference between the pressure coefficients for the two structurally-similar polymers is within the experimental uncertainty.

**Volume effects on dynamics.** The use of Arrhenius plots (Figure 6) and activation volumes begs the question of what governs the relaxation times in POB. Obviously the fact that an isothermal pressure changes induces a change in $\tau$ demonstrates that thermal energy alone is not the dominant control parameter. To assess directly the dependence of the relaxation times on volume, we measured the variation of the specific volume, V, with temperature and pressure (Figure 11). All measurements were carried out in the equilibrium liquid. Using the Tait form for the equation of state[89], we obtain for POB above $T_g$

$$V(T,P) = (0.9530 + 6.45 \times 10^{-4}T + 5.7 \times 10^{-7}T^2)\left[1 - 0.0894\ln\left(1 + P/171\exp(-0.00513T)\right)\right] \quad (6)$$

in units of ml/g, with T and P in Celsius and MPa (the former by convention). Using eq. 7, we calculate the specific volume corresponding to each measurement condition, and plot in Figure 12 the relaxation times as a function of V. They are not a unique function of V, however, since temperature (thermal energy) exerts some influence.

We recently proposed a method to quantify the effects of temperature and volume on $\tau_\alpha$, whereby segmental relaxation times obtained under various conditions of T and P collapse onto a single master curve[45,46,90]. The superpositioning is based on expressing $\tau_\alpha$ as a function of the product $TV^\gamma$, where $\gamma$ is a material-specific constant. More recently, we extended this $TV^\gamma$-scaling to the normal mode relaxation times for PPG and PI [91]. Most intriguing was the finding that the $\gamma$ required to superpose the normal mode relaxation times was the same that for $\tau_\alpha$, notwithstanding that for these low molecular weight polymers, the two modes differ in both temperature- and pressure-dependences. Such results are congruent with the idea that segmental and chain motions are both governed by the same local (monomeric) friction coefficient; that is, the variation in relaxation times with temperature or pressure is mainly driven by the T- and P-dependences of this friction coefficient.

Although the P-dependences for the normal and segmental modes for POB are quite similar (Table 2), the respective relaxation times exhibit different dependences on T near $T_g$ (Figure 6 and Figure 7 insert). Thus, it is of interest to ascertain if master curves of the $\tau$ can be obtained, and if so, whether $\gamma$ is the same for the two modes. In Figure 13 all relaxation times measured herein are plotted versus $T^{-1}V^{-\gamma}$ (we use the reciprocals to obtain a scaled Arrhenius plot). The exponent is adjusted to obtain the superpositioning seen in the figure, which corresponds to a value of $\gamma = 2.65$, for both $\tau_\alpha$ and $\tau_N$. This is a fairly small value, reflecting the relatively strong influence of thermal energy, as opposed to volume, in governing the T-dependence of $\tau_\alpha$.

A relationship exists between $\gamma$ and the ratio of the isochoric and isobaric activation enthalpies, $E_V/E_P$[45]

$$\frac{E_V}{E_P} = (1 + T_g \alpha_P \gamma)^{-1} \qquad (7)$$

where $\alpha_P$ is the thermal expansion coefficient. This ratio is of interest, because it provides a direct measure of the degree to which the changes in $\tau$ with temperature are due to the accompanying volume changes, as opposed to changes in thermal energy. For P = 0.1 MPa and using $\alpha_P(T_g) = 6.170\times10^{-4}$ K$^{-1}$ as determined from the PVT data for POB, we calculate $E_V/E_P = 0.75$ at $T_g$. This is comparable to the value for a PPG of similar molecular weight[92], but larger than the $E_V/E_P$ ratios found for non-associated glass-formers[93]. The presence of hydrogen bonds increases the role of thermal energy relative to that of the specific volume.

In Figure 13 we have included the conductivities measured under the various conditions of T and P. Given the decoupling of the relaxation times and σ, we do not expect the latter to exhibit the same scaling exponent. This is indeed borne out in the figure; a smaller value of $\gamma \sim 1.7$ gives the best superpositioning of the conductivity data. This is unsurprising, since conductivity results from translation of ions, rather than reorientation of polymer segments, and thus samples different local environments. Note that the superpositioning of the σ(T,P) is only approximate, certainly having more scatter than the master curves for either $\tau_\alpha$ or $\tau_N$. Moreover, we point out that a value of $\gamma = 1.7$ is too small to justify neglect of the attractive part of the intermolecular potential.

**Secondary relaxations.** There are two other notable features in the dielectric spectra of POB, seen primarily at lower temperatures and higher pressures. In Figure 14, the dielectric loss is shown at different temperatures for three pressures. There is a deviation from the fit of the segmental relation to eq. 2, beginning about two decades higher frequency than the α-peak. This is followed at still higher frequencies by a steep rise, suggestive of the onset of another dispersion. Of course, secondary relaxations are well known in polymers[94]. Of particular interest, because of its connection to the glass transition, is the Johari-Goldstein (JG) secondary relaxation[1,2]. Involving all atoms of the repeat unit, the JG process is the fastest secondary relaxation, and when in close proximity to the α-process, can appear as an "excess wing" (that is, rather than a distinct peak, the JG is manifested as dielectric absorption above the fitted curve on the high frequency flank of the α-relaxation). Interestingly, the JG process has been found to have a pressure dependence closely related to that of the α-process[2,3,95], consistent with the idea that the JG may serve as the precursor to the glass transition. There is a relationship between the $\tau_\alpha$ and $\tau_{JG}$ [59,96]

$$\tau_{JG} = t_c^{1-\beta_{KWW}} \tau_\alpha^{\beta_{KWW}} \tag{8}$$

with $t_c \sim 2\times10^{-12}$ s. Using the value of $\beta_{KWW} = 0.51$, the JG peak is calculated to lie about six decades higher in frequency than the α-peak (as indicated by the arrows in Fig. 14). This falls well below the high frequency peak

For hydrogen bonded glass-formers, an additional secondary peak beyond the JG process is often observed, for example in sorbitol[16] and PPG[3,95]. In contrast with the JG, this high frequency peak has an almost negligible pressure dependence. This appears to be the case for the high frequency secondary peak in POB. As seen in Figure 15, while the α-process slows by many decades, the position of the high frequency dispersion is almost unchanged. Unfortunately, the range of conditions over which this peak could be observed is too limited to allow quantitative analysis of its properties.

**Concluding Remarks**

Dielectric measurements on POB at different temperatures and pressures reveal the presence of five dynamic processes. Progressing from lower to higher frequency, these include conductivity of mobile ionic impurities, the normal mode relaxation involving global chain motions, the α-process reflecting local segmental motion, an excess wing which can be identified with the Johari-Goldstein relaxation, and a high frequency secondary relaxation commonly found in hydrogen-bonded materials. Only the normal and segmental modes have dispersions in the dielectric loss falling within the range of measurement conditions herein. When compared at a fixed value of the relaxation time, the shape of either peak is constant, independent of T and P.

As commonly found for low molecular weight polymers, the relaxation times for the normal and segmental processes exhibit differences in T-dependences near $T_g$. Similar differences, although of smaller amplitude, are also observed by approaching $T_g$ by increasing pressure. However, in the accessible dynamic range, we find that for POB, hydrostatic pressure exerts almost an equivalent effect on $\tau_N$ and $\tau_\alpha$. Ostensibly, this contradicts previously reported results for PPG and PI [12,14]; however, a re-analysis of the data for PPG and PI reveals similar behavior, when the comparison of pressure-sensitivities is made over the same range of pressure (rather than same range of τ).

The relaxation times for both relaxation modes are not uniquely determined by either T or V, although from the ratio of the isochoric and isobaric activation enthalpies, $E_V/E_P = 0.75$, we conclude that $\tau_\alpha$ is governed more strongly by changes in temperature than by the accompanying volume changes. Generally, relaxation times for polymers are less affected by density in comparison to $\tau_\alpha$ for small-molecule glass-formers. The reason for this is the same reason the density-normalized dielectric strength of the normal mode is almost invariant to pressure (isotherms in Fig. 3). To wit, pressure has negligible effect on the chain dimensions, so that interactions among directly bonded segments are unaffected by pressure. This is not the case for molecular liquids, wherein every near-neighbor interaction is intermolecular, and these are all sensitive to density.

The $\tau_N$ and $\tau_\alpha$ data can be superposed to form single respective curves when expressed as a function of $TV^{2.65}$. This equivalence of the scaling-exponent for the normal and segmental modes was found previously for PPG and PI, and is consistent with the idea that there is a single monomeric friction coefficient, underlying both the local and global dynamics.

The ionic conductivity, however, does not superpose when plotted versus $TV^{2.65}$. The conductivity also shows the customary enhancement, relative to the α-relaxation, as $T_g$ is approached by either cooling or increasing pressure. It is interesting that the magnitude of the enhancement is the same relative to either $\tau_\alpha$ or $\tau_N$. Translational motions changing more with T than do local reorientational modes has been ascribed to spatially heterogeneous dynamics[86]. The length scale for the dynamic heterogeneity of the glass transition has been measured to be in the range from 1 nm to 4 nm[97,98,99,100]. We can estimate the radius of gyration of the POB to be ~ 2.3 nm. Since we do not observe the enhancement of the normal mode motion, as seen for the conductivity, it is tempting to conclude that the heterogeneity length scale in POB must be > 2.3 nm, whereby it is averaged out. On the other hand, there is an alternative explanation for the enhanced translation mobility that does not invoke dynamic heterogeneity[87].

Eq. 5 is drawn from the idea that the JG relaxation serves as a precursor to the α-process[2,59,95,96], and accordingly, it is of central importance to understanding the dynamics of the glass transition. In POB, the JG relaxation is barely resolved, appearing as an excess-wing on the high frequency flank of the α-peak. An additional high frequency secondary peak is evident in the POB spectra, which appears to be relatively insensible to pressure, although it lies at too high frequency to be well-characterized.

Finally, it is of interest to compare POB to the structurally similar PPG. The latter differs from POB by having a smaller pendant moiety at the second carbon atom (methyl group in place of ethyl). This minor change results in significant changes in the dielectric strength of the normal mode, attributed to differences of the mean square end-to-end distance of the chains. In Table 3 we compare the properties determined herein for POB to those for PPG having an equivalent degree of polymerization. The different pendant group does not modify the dynamics substantially, although POB is less fragile and less influenced by volume changes (i.e., larger $E_V/E_P$). Nevertheless, for both polymers, the scaling exponent γ is small, implying that volume and congested dynamics exert a minor role on the thermally-activated dynamics.

**Acknowledgements**

We thank Colin Booth of the University of Manchester for graciously providing the POB sample. This work was supported by the Office of Naval Research.

**References**

1   Johari, G.P.; Goldstein, M.; *J.Chem.Phys.* **1970**, *53*, 2372.
2   Ngai, K.L.; Paluch, M. *J.Chem.Phys.* **2004**, *120*, 857.

3    Casalini, R.; Roland, C.M. *Phys. Rev. Lett.* **2003**, *91* 015702.

4    Roland, C.M.; Schroeder, M.J.; Fontanella, J.J.; Ngai, K.L. *Macromolecules* **2004**, *37*, 2630.

5    Casalini, R.; Ngai, K.L.; Roland, C.M. *Journal Of Chemical Physics*, **2000,** *112*, 5181.

6    Ngai, K.L.; Casalini, R. *Physical Review B,* **2002**, *66*, 132205.

7    Rivera, A.; Leon, A.; Varsamis, C.P.E.; Chryssikos, A.D.; Ngai, K.L.; Roland, C.M.; Buckley, L.J. *Physical Review Letters* **2002**, *88*, 125902.

8    Kisliuk, A.; Novikov, V.N.; Sokolov, A.P. *Journal Of Polymer Science Part B-Polymer Physics* **2002**, *40*, 201.

9    Angell C.A., Wang L.M., Mossa S.; Yue Y.Z., Copley J.R.D. *AIP Conference proceedings* **2004**, *78*, 473.

10   Williams, G. *Trans. Faraday Soc.* **1965**, *61*, 1564.

11   Bendler, J.T.; Fontanella, J.J.; Schlesinger, M. *Phys. Rev. Lett.* **2001**, *87*, 195503.

12   Floudas, G.; Reisinger, T. *J. Chem. Phys.* **1999**, *111*, 5201.

13   Mierzwa, M.; Floudas, G.; Dorgan, J.; KNauss, D.; Wegner, J. *J. Non-Cryst. Solids* **2002**, *307-310*, 296.

14   Roland, C.M.; Psurek, T.; Pawlus, S.; Paluch, M. *J. Polym. Sci. Polym. Phys. Ed.* **2003**, *41*, 3047.

15   Andersson, S.P.; Andersson, O. *Macromolecules* **1998**, *31*, 2999.

16   Hensel-Bielowka, S.; Paluch, M.; Ziolo, J.; Roland, C.M. *J. Phys. Chem. B* **2002**, *106*, 12459.

17   Casalini, R.; Capaccioli, S.; Lucchesi, M.; Rolla, PA.; Corezzi, S. *Physical Review E* **2001**, *63*, 031207.

18   Serghei, A.; Kremer, F. *Physical Review Letters* **2003**, *91*, 165702.

19   Schönhals, A.; Goering, H.; Schick, C. *Journal Of Non-Crystalline Solids* **2002**, *305*, 140.

20   Jackson, C.L.; McKenna, G.B. *Chemistry Of Materials* **1996**, *8*, 2128.

21   Simon, S.L.; Park, J.Y.; McKenna, G.B. *European Physical Journal E* **2002**, *8*, 209.

22   Mijovic J., Han Y.F., Sun M.Y., Pejanovic S. *Macromolecules* **2003**, *36*, 4589.

23   Andjelic, S.; Mijovic, J. *Macromolecules* **1998**, *31* 8463.

24   Kramarenko, V.Y.; Ezquerra, T.A.; Sics, I.; Balta-Calleja, F.J.; Privalko, V.P. *Journal Of Chemical Physics* **2000**, *113*, 447.

25   Tombari, E.; Ferrari, C.; Salvetti, G.; Johari, G.P. *Journal Of Physics-Condensed Matter,* **1997***, 9,* 7017.

26   Casalini, R.; Livi, A.; Rolla, PA.; Levita, G.; Fioretto, D. *Physical Review B* **1996**, *53*, 564.

27   Casalini, R.; Corezzi, S.; Fioretto, D.; Livi, A.; Rolla, PA. *Chemical Physics Letters* **1996**, *258*, 470.

(28) Williams, G.; Smith, I.K.; Aldridge, G.A.; Holmes, P.A.; Varma, S. *Macromolecules* **2001**, *34*, 7197.

(29) Urakawa, O.; Swallen, S. F.; Ediger, M. D.; von Meerwall, E. D. *Macromolecules* **2004**, *37*, 1558

(30) Roland, C.M.; Ngai, K.L.; Plazek, D.J. *Macromolecules* **2004**, *37*, 7051.

(31) Santangelo, P.G.; Roland, C.M. *Macromolecules* **1998**, *31*, 3715.

(32) Stockmayer, W.H. *Pure Appl.Chem.* **1967**, *15*, 539.

(33) Lunkenheimer, P.; Schneider, U.; Brand, R.; Loidl, A. *Contemporary Physics* **2000**, *41*, 15.

(34) Ferry, J.D. *Viscoelastic Properties of Polymers*, 3rd edition (Wiley, New York, 1980).

(35) Cohen, M.H.; Grest, G.S. *Phys.Rev.B* **1979**, *20*, 1077.

(36) Turnbull, D.; Cohen, M. *J. Chem. Phys.* **1970**, *52*, 3038.

(37) Goldstein, M. *J.Chem.Phys.* **1969**, *51*, 3728.

(38) Adam, G.; Gibbs, J.H. *Journal of Chemical Physics* **1965**, *43*, 139.

(39) Ngai, K.L.; Tsang, K.Y. *Physical Review E* **1999**, *60*, 4511.

(40) Ngai, K.L. *Journal of Physics-Condensed Matter* **2003**, *15*, S1107.

(41) Ferrer, M.L.; Lawrence, Ch.; Demirjian, B.G.; Kivelson, D.; Alba-Simonesco, Ch.; Tarjus, G. *J. Chem. Phys.* **1998**, *109*, 8010.

(42) Williams, G. in *Dielectric Spectroscopy of Polymeric Materials*; Runt, J.P.; Fitzgerald, J.J. Eds.; ACS, Washington DC, 1997.

(43) Paluch, M.; Casalini, R.; Roland, C.M.; *Phys. Rev. B* **2002**, *66*, 092202.

(44) Paluch, M.; Casalini, R.; Roland, C.M.; Meier, G.; Patkowski, A. *J. Chem. Phys.*, **2003**, *118*, 4578.

(45) Casalini, R.; Roland, C.M. *Phys. Rev. E* **2004**, *69*, 062501.

(46) Casalini, R.; Roland, C.M. *Col. Poly. Sci.* **2004**, *283*, 107.

(47) Dreyfus, C.; Aouadi, A.; Gapinski, J.; Matos-Lopes, M.; Steffen, W. Patkowski, A. Pick, R.M. *Physical Review E* **2003**, *68*, 011204.

(48) Tölle, A. *Reports on Progress in Physics* **2001**, *64*, 1473.

(49) Shell, M.S.; Debenedetti, P.G.; La Nave, E.; Sciortino, F. *Journal of Chemical Physics* **2003**, *118*, 8821.

(50) Weeks, J.D.; Chandler, D.; Andersen, H.C. *Journal of Chemical Physics* **1971**, *54*, 5237.

(51) Widom, B. *Science* **1967**, *157*, 375.

(52) Longueth-Higgins, H.C.; Widom, B. *Mol. Phys.* **1964**, *8*, 549.

(53) Hoover, W.G.; Ross, M. *Contemporary Physics* **1971**, *12*, 339.

54 March, N.H.; Tosi, M.P. *Introduction to Liquid State Physics*; World Scientific: Singapore, 2002.

55 Hall, C.K.; Helfand, E.J. *J. Chem. Phys.* **1982**, *77*, 3275.

56 Doi, M.; Edwards, S. F. *The Theory of Polymer Dynamics*; Clarendon: Oxford, 1986.

57 Schönhals, A.; Kremer, F.; Schlosser, E. *Progr. Colloid. Polym. Sci.* **1993**, *91*, 39.

58 Schönhals, A.; Stauga, R. *J. Non-Cryst. Sol.* **1998**, *235-237*, 450.

59 León, C.; Ngai, K.L.; Roland, C.M. *J. Chem. Phys.* **1999**, *110*, 11585.

60 Mattsson, J.; Bergman, R.; Jacobsson, P.; Börjesson, L. *Phys. Rev. Lett.* **2003**, *90*, 075702.

61 Baur, M.E.; Stockmayer, W.H. *J. Chem. Phys.* **1965**, *43*, 4319.

62 Fontanella, J.J.; Wintersgill, M.C.; Smith, M.K.; Semancik, J.; Andeen, C.G. *J. Appl. Phys.* **1986**, *60*, 2665.

63 Casalini, R.; Roland, C.M.; *Phys. Rev. Lett.* **2003**, *91*, 15702.

64 Wang, C.H.; Fytas, G.; Lilge, D.; Dorfmuller, Th. *Macromolecules* **1981**, *14*, 1363.

65 Wang, C.H.; Gong, S.S.; *J. Chem. Phys.* **2003** *119*, 11445.

66 Cochrane, J.; Harrison, G.; Lamb, J.; Phillips, D.W. *Polymer* **1980**, *21*, 837.

67 Barlow, A.J.; Erginsav, A. *Polymer* **1975**, *16*, 110.

68 Kyritsis, A.; Pissis, P.; Mai, S.M.; Booth, C. *Macromolecules* **2000**, *32*, 4581.

69 Watanabe, H. *Prog.Polym.Sci.* **1999**, *24*, 1253.

70 Adachi, K.; Kotaka, T. *Prog.Polym.Sci.* **1993**, *18*, 585.

71 Ngai, K.L.; Roland, C.M. *Macromolecules* **1993**, *26*, 6824.

72 Adachi, K. *Dielectric Spectroscopy of Polymeric Materials*, Ed.J.P.Runt, J.J.Fitzgerald, ACS, Washington DC, 1997.

73 *Broadband Dielectric Spectroscopy*, Kremer, F.; Schonhals, A., eds.; Springer: New York, 2002.

74 *Dipole Moments and Birefringence of Polymers*, Riande, E.; Saiz, E., eds.; Prentice Hall: Englewood Cliffs, NJ, 1992.

75 Alvarez, F.; Alegria, A.; Colmenero, J. *Phys. Rev. B* **1991**, *44*, 7306.

76 Böhmer, R.; Ngai, K.L.; Angell, C.A.; Plazek, D.J. *J.Chem.Phys.* **1993**, *99*, 4201.

77 Plazek, D.J.; Schlossser, E.; Schonhals, A.; Ngai, K.L. *J. Chem. Phys.* **1993**, *98*, 6488.

78 Mijovic, J.; Sun, M.; Han, Y. *Macromolecules* **2002**, *35*, 6417.

79 Schönhals, A. *Macromolecules* **1993**, *26*, 1309.

80 Mierzwa, M.; Floudas, G.; Dorgan, J.; KNauss, D.; Wegner, J. *J. Non-Cryst. Solids* **2002**, *307-310*, 296.

81 Floudas, G.; Gravalides, C.; Reisinger, T.; Wegner, G. *J.Chem.Phys.* **1999**, *111*, 9847.

[82] Rössler, E. *Phys. Rev. Lett.* **1990**, *65*, 1595.

[83] Fujara, F.; Geil, H.; Sillescu, H.; Fleischer, G.Z. *Z. Phys. B* **1992**, *88*, 195.

[84] Cicerone, M.T.; Blackburn, F.R.; Ediger, M.D. *Macromolecules* **1995**, *28*, 8224.

[85] Corezzi, S.; Campani, E.; Rolla, P.A.; Capaccioli, S.; Fioretto, D. *J. Chem. Phys.* **1999**, *111*, 9343.

[86] Cicerone, M. T.; Wagner, P. A.; Ediger, M. D. *J. Phys. Chem. B* **1997**, *101*, 8727.

[87] Ngai, K. L. *J. Phys. Chem. B* **1999**, *103*, 10684.

[88] Atake, T.; Angell, C.A. *J. Phys. Chem.* **1979**, *83*, 3218.

[89] Tait, P.G. *Physics and Chemistry of the Voyage of H. M. S. Challenger* HMSO, London, **1888**, Vol. 2, Part 4

[90] Casalini, R.; Roland, C.M. *Phys.Rev. B, in press (arxiv.org/abs/cond-mat/0412258).*

[91] Roland, C.M.; Casalini, R.; Paluch, M. *J. Polym. Sci. Polym. Phys. Ed.* **2004**, *42*, 4313.

[92] Casalini, R.; Roland, C.M. *J. Chem. Phys.* **2003**, *119*, 11951.

[93] Roland, C.M.; Paluch, M.; Pakula, T.; Casalini, R. *Phil. Mag. B* **2004**, *84*, 1573.

[94] McCrum, N. G.; Read, B. E.; Williams, G. *Anelastic and Dielectric Effects in Polymeric Solids*; Dover: New York, 1991.

[95] Casalini, R.; Roland, C.M. *Phys. Rev. B* **2004**, *69*, 094202.

[96] Ngai, K.L. *Phys. Rev. E* **57**, 7346, (1998).

[97] Tracht, U.; Wilhelm, M.; Heuer, A.; Feng, H.; Schmidt-Rohr, K.; Spiess, H.W. *Phys. Rev. Lett.* **1998**, *81*, 2727.

[98] Reinsberg, S.A.; Qiu, X.H.; Wilhelm, M.; Spiess, H.W.; Ediger, M.D. *J. Chem. Phys.* **2001**, *114*, 7299.

[99] Hempel, E.; Hempel, G.; Hensel, A.; Schick, C.; Donth, E. *J. Phys. Chem. B* **2000**, 104, 2460

[100] Vyazovkin, S.; Dranca, I. *J. Phys. Chem. B* **2004**, *108*, 11981.

Table 1. VFTH Parameters (P=0.1 MPa)

|  | $\tau_N$ | $\tau_\alpha$ |
|---|---|---|
| log ($\tau_\infty$ /s) | -9.86 ± 0.33 | -12.53 ± 0.43 |
| B | 1276 ± 99 K | 1249 ± 112 K |
| $T_0$ (K) | 158.8 ± 2.4 | 161.7 ± 2.3 K |

Table 2. Activation Volumes

| T (K) | P (MPa) | ΔV (ml/mol) | |
|---|---|---|---|
| | | normal mode | segmental mode |
| 246.6 K | 0-140 | 69 ± 1 | 69 ± 1 |
| 273.2 K | 0-190 | 52 ± 3 | 52 ± 3 |
| 297.4 K | 180-350 | 39 ± 1 | 40 ± 1 |

Table 3. Comparison of Relaxation Properties of POB and PPG

|  | POB | PPG |
|---|---|---|
| $T_g$ (K) | 199.0 | 198 [59] |
| $dT_g/dP$ (K/GPa) | 155 ± 25 | 177 ± 36 [14] |
| fragility | 78 ± 1 | 83 ± 2 [59] |
| $\beta_{KWW}$ | 0.51 ± 0.01 | 0.55 [59] |
| $\gamma$ | 2.65 ± 0.05 | 2.5 ± 0.35 [91] |
| $E_V/E_P$ | 0.75 ± 0.01 | 0.66 ± 0.03 [95] |
| $d\Delta\varepsilon_\alpha/dP$ (GPa$^{-1}$) | ~ 1 | |
| $d\Delta\varepsilon_N/dP$ (GPa$^{-1}$) | ~ 0 [14] | |

**Figure Captions**

Figure 1. Comparison of dielectric loss spectra of POB and PPG having the same number of repeat units (= 67), measured at similar temperatures and values of the relaxation times. The peak frequencies of segmental relaxation were $f_{max}$= 98.5 kHz for POB and $f_{max}$ =71.2 kHz for PPG. The POB spectra is shown as measured (○; T=297.4 K, P = 351 MPa)), and after multiplying by a factor 1.45 to superimpose the two α-peaks (△). The PPG spectra (■), taken from ref. 14, is for T = 293 K and P = 616 MPa.

Figure 2. Representative dielectric loss spectra at two elevated pressures, showing the fitted contributions to the ionic conductivity (dashed-dotted line), the normal mode (dashed line) and the segmental peak (dotted line). There is an additional peak, with a maximum at frequencies beyond the measured range.

Figure 3. Dielectric strength, normalized by the mass density, of the normal and α-modes plotted as a function of the normal mode relaxation time for an isobar and three isotherms. The relative magnitude of Δε/ρ for the segmental mode increases substantially with increasing τ. At fixed τ, increasing pressure increases the dielectric strength of the segmental model, while having the opposite effect on Δε/ρ for the normal mode. Solid lines, representing average values, are drawn for the normal mode isotherms, to indicate their near invariance to τ.

Figure 4. Comparison of the peaks for the normal and segmental modes at an approximately constant value of $\tau_\alpha \sim 10^{-3}$ s; the ordinate has been normalized by the peak intensity of the segmental mode. The relative intensity of the normal mode increases with increase of either pressure or temperature. Note that there is negligible change in either the peak shapes or the separation of the two peaks.

Figure 5. Havriliak-Negami shape factors (eq. 1) plotted versus the normal-mode relaxation times: P = 0.1 MPa (■●); T = 246.6 K (◨◓), 273.2 K (□○), and 297.4 K (◨◓).

Figure 6. Normal mode (□) and segmental (○) relaxation times measured at ambient pressure. The crosses represent the latter shifted by 2.6 decades, to superimpose at higher temperatures on the $\tau_N$. The solid lines are fits to the VFTH equation, with log $\tau_\infty$ (s) = -9.86 ± 0.33, b = 1276 ± 99 K , and $T_0$ = 158.8 ± 2.4 K for $\tau_N$, and log $\tau_\infty$ (s) = -12.62 ± 0.34, b = 1269 ± 92 K, and $T_0$ = 161.3 ± 1.9 K for $\tau_\alpha$. The dashed lines represent the experimental results for POB M = 5,085 D reported by Kyrjitsis et al.[68]. The inset is a fragility plot of the segmental relaxation times, having a slope at $T_g$ equal to 77.

Figure 7. Segmental relaxation times as a function of normal mode relaxation times for three isotherms. The pressure variation was in the range from 11 to 952 MPa. The slope of the fitted line reveals the near equivalence of the pressure dependences for the two relaxation modes. The inset shows the logarithm of the ratio of the two relaxation times, plotted versus the normal mode relaxation time (symbols same as in main figure).

Figure 8. Segmental relaxation times as a function of normal mode relaxation times for six isotherms measure for PPG (M = 4000 D); data from ref. 14. The slope was obtained by simultaneously fitting all data.

Figure 9 Segmental relaxation times as a function of normal mode relaxation times for PI of respective molecular weight equal to 1,200 D (hollow symbols) [81] and 10,200 D (filled symbols) [12].

Figure 10. Ionic conductivity versus segmental relaxation time, yielding log (σ/$\tau_\alpha$) = -0.92 for P = 0.1 MPa (●), = -0.89 for T = 246.6 K (◐), = -0.84 for T = 273.2 K (○), and = -0.78 for T = 297.4 K (◑).

Figure 11 Specific volume for POB as function of T at various P (ranging from 0.1 to 200 MPa in steps of 10 MPa). The solid lines represent the fit to the Tait equation of state, with the parameters given in the text.

Figure 12. Segmental (upper panel) and normal mode (lower) relaxation times as a function of the specific volume. The symbols are the same as in Figure 5.

Figure 13. Dielectric normal mode (squares) and segmental (circles) relaxation times measured as a function of temperature at P = 0.1 MPa (■●), and as a function of pressure at T = 246.6 K (◨◐), 273.2 K (□○), and 297.4 K (◧◑). The data were superposed using an exponent γ = 2.65. Ionic conductivities, for which γ = 1.7, superpose poorly.

Figure 14. Representative dielectric α-peaks measured at atmospheric pressure and T = 197.0K (□), at 416.7 MPa and 246.6 K (◐), and at 684.5 MPa and 273.2 K (○), showing the high frequency deviation from eq. 1 ("excess wing"), as well as the emergence of an additional secondary process at still higher frequencies. The arrows denote the calculated position (eq. 8) of the JG process for the two elevated pressure spectra.

Figure 15. Spectra at T= 246.6K and various pressures (as indicated in MPa). The high frequency rise in the dielectric loss at high frequency, suggestive of a secondary relaxation (with $f_{peak} > 10^6$ Hz), is relatively insensitivity to pressure.

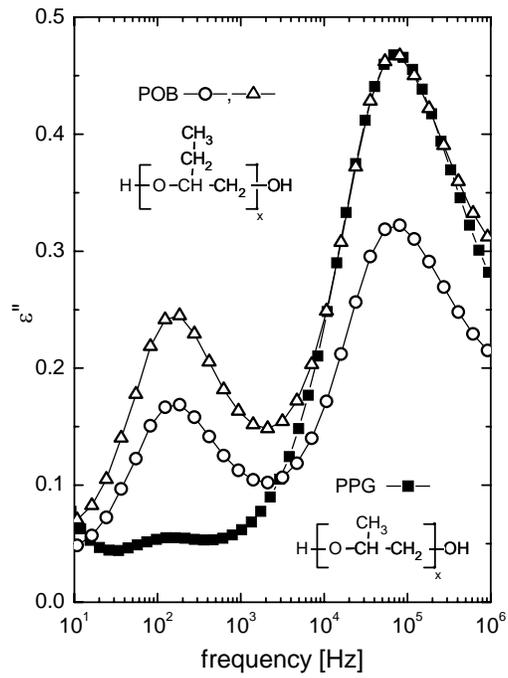

Figure 1

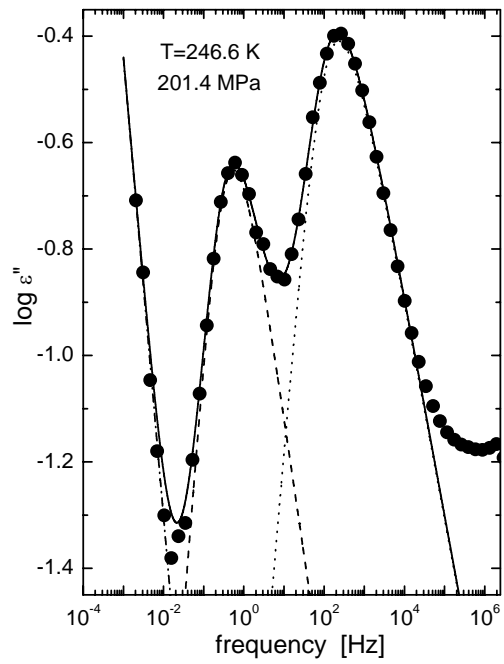

Figure 2

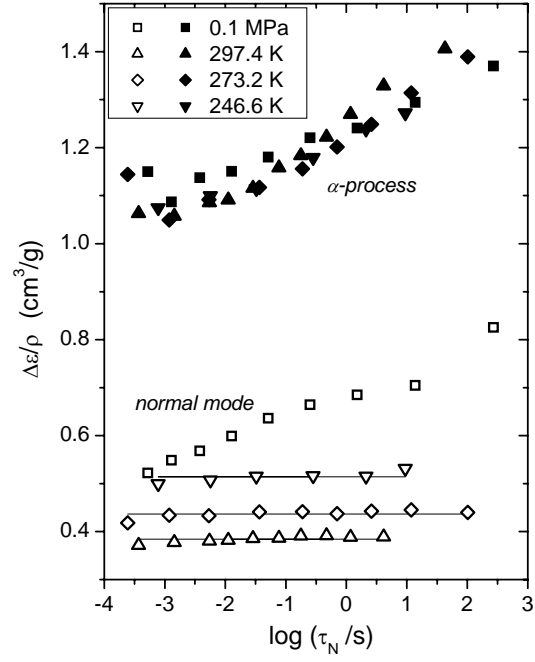

Figure 3

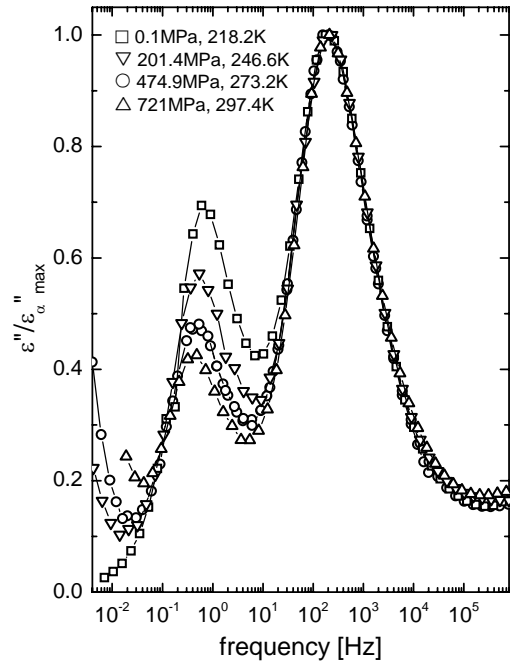

Figure 4

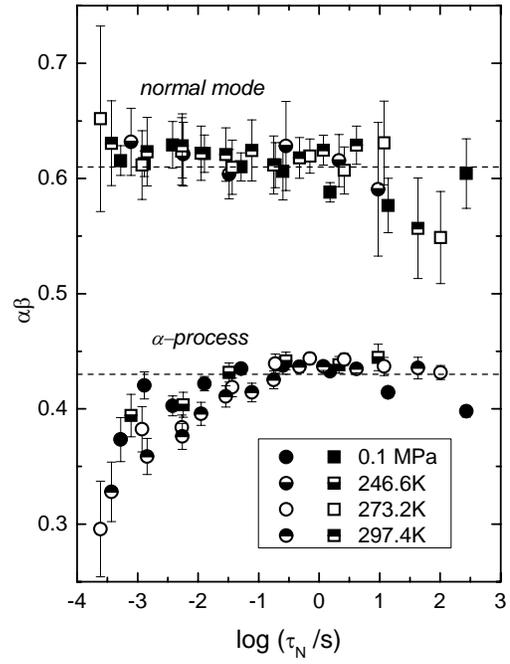

Figure 5

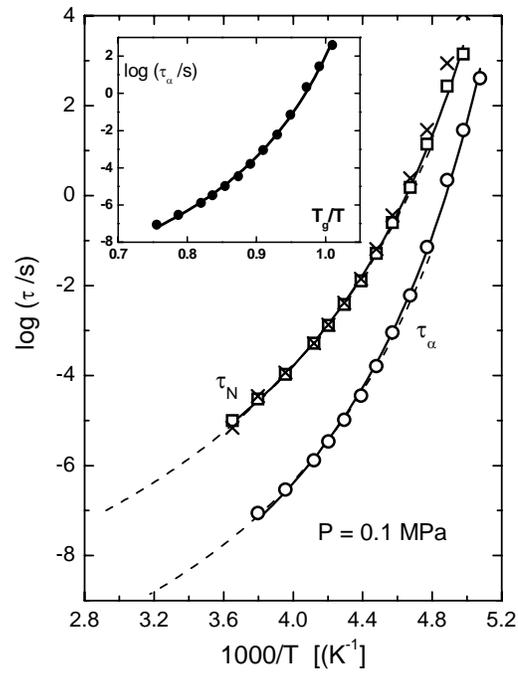

Figure 6

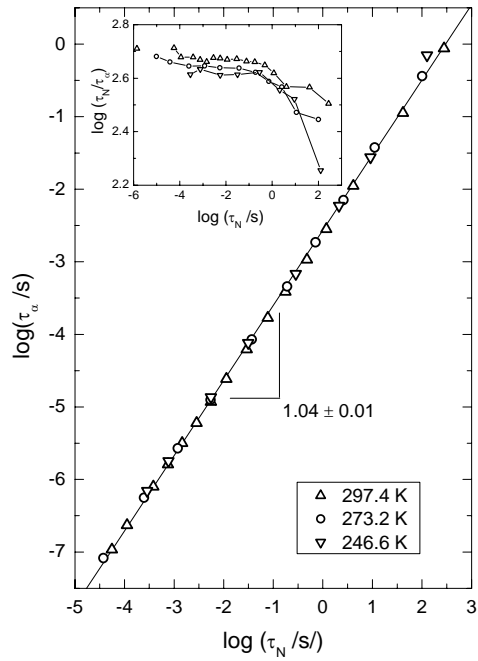

Figure 7

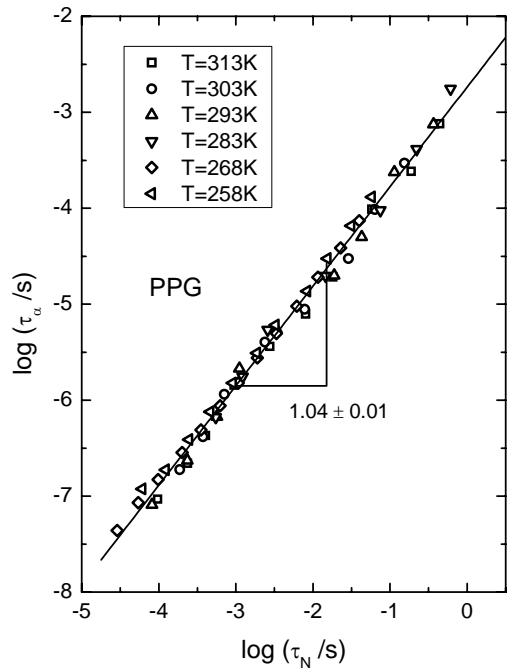

Figure 8

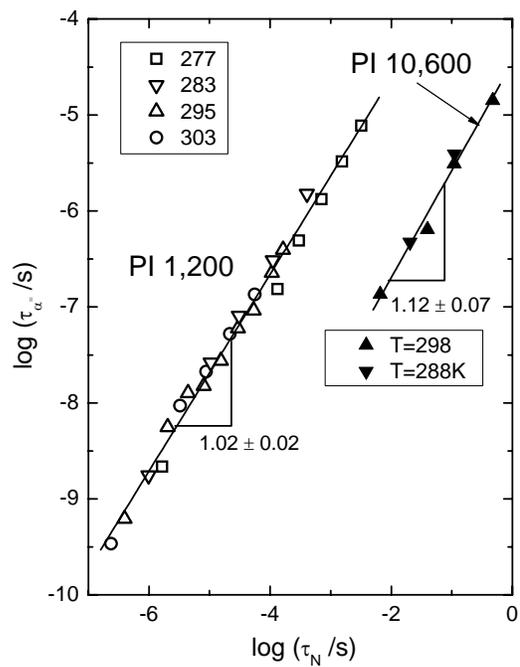

Figure 9

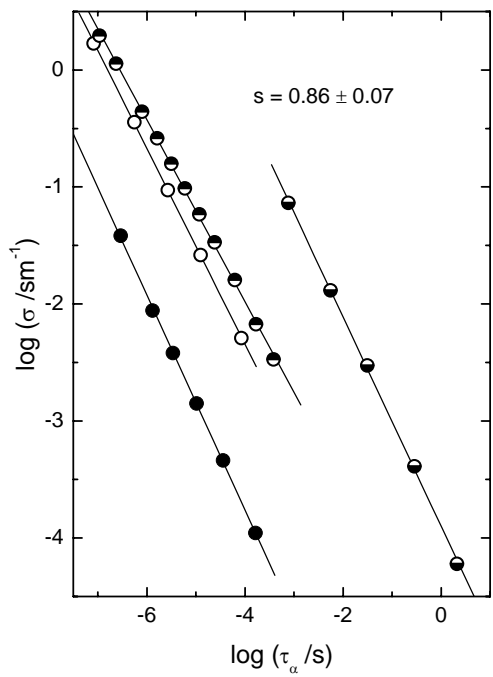

Figure 10

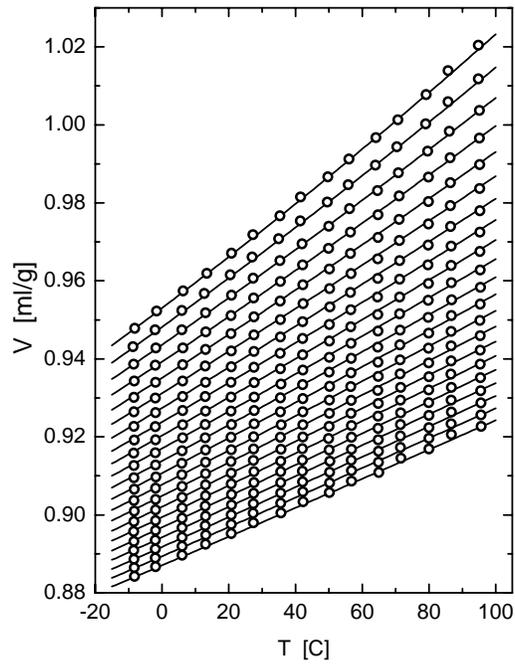

Figure 11

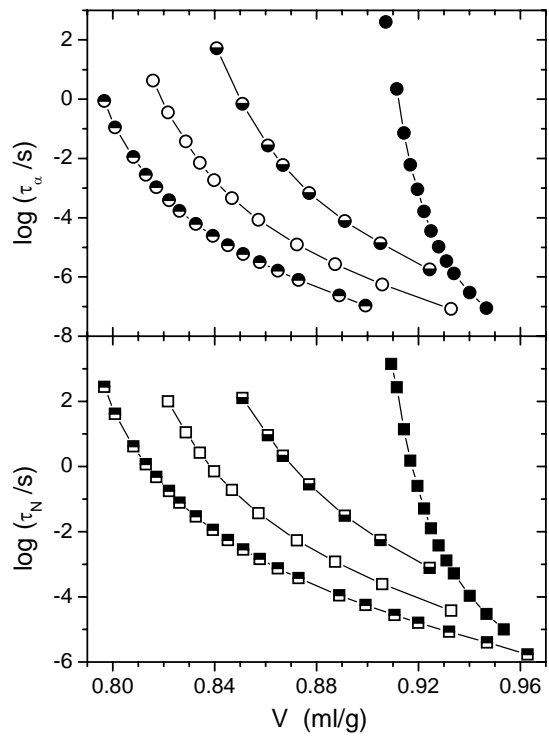

Figure 12

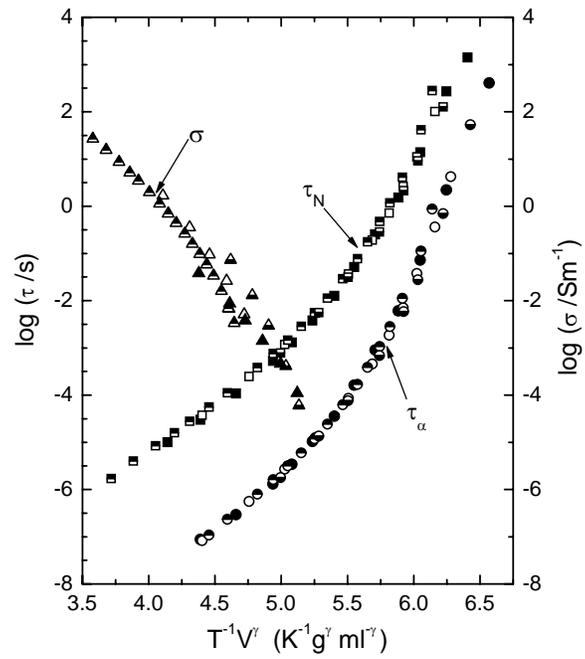

Figure 13

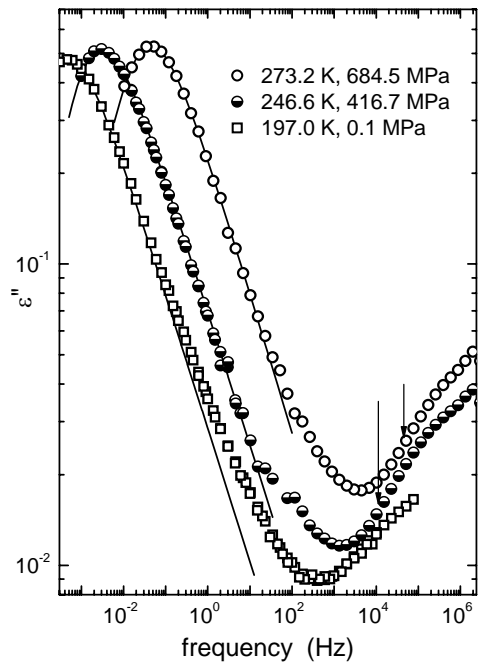

Figure 14

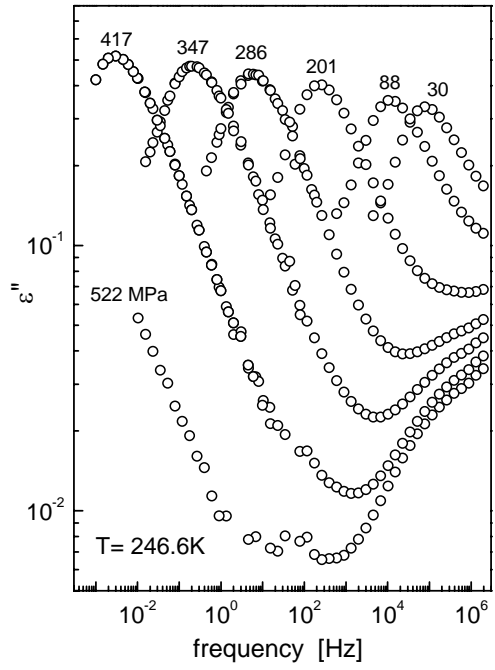

Figure 15